\documentclass[pra,onecolumn,floatfix,superscriptaddress,longbibliography,notitlepage, nofootinbib]{revtex4-1}
\usepackage[utf8]{inputenc}
\usepackage{braket}
\usepackage{amsmath}

\DeclareMathOperator{\Tr}{Tr}
\usepackage{dcolumn}   % needed for some tables
\usepackage{bm}        % for math
\usepackage{amssymb}
\usepackage{mathtools}
\usepackage{tikz}
\usepackage{qcircuit}
\usepackage{appendix}
\usepackage{hyperref}

\usepackage{subcaption}
\usepackage{amsmath,amsfonts,amssymb}
\usepackage{mathtools}
\usepackage{thmtools,thm-restate}
\usepackage{algorithm}
\usepackage{mathrsfs}
\usepackage{tikz}
\usepackage{subcaption}
\usepackage{pgfplots}
\usetikzlibrary{3d}
\usepackage{tkz-euclide}
\usepackage{mathtools}
\newcommand{\xbf}{\textbf{x}}

\newcommand{\Ibb}{\mathbb{I}}

\usepackage{algpseudocode}

\newtheorem{definition}{Definition}
\newtheorem{theorem}{Theorem}
\newtheorem{lemma}{Lemma}

\newcommand{\revise}[1]{\textcolor{black}{#1}}

\usetikzlibrary{arrows.meta}

\def\BibTeX{{\rm B\kern-.05em{\sc i\kern-.025em b}\kern-.08em
    T\kern-.1667em\lower.7ex\hbox{E}\kern-.125emX}}

\definecolor{ao}{rgb}{0.0, 0.5, 0.0}

\newif\ifshowcorrections
\showcorrectionsfalse
\DeclareRobustCommand{\corr}[1]{\ifshowcorrections\textcolor{blue}{#1}\else#1\fi}
\newenvironment{corrblock}{\ifshowcorrections\color{blue}\fi}{}
    
\begin{document}
%\title{Improved Quantum Power Method and Numerical Integration Using Quantum Singular Value Transformation}
\title{Quantum Power Method and Numerical Integration Using Quantum Singular Value Transformation} % [E7]
\author{Nhat A. Nghiem}
\email{nhatanh.nghiemvu@stonybrook.edu}
\affiliation{Department of Physics and Astronomy, State University of New York at Stony Brook, Stony Brook, NY 11794-3800, USA}
\affiliation{C. N. Yang Institute for Theoretical Physics, State University of New York at Stony Brook, Stony Brook, NY 11794-3840, USA}
\author{Hiroki Sukeno}
\affiliation{Department of Physics and Astronomy, State University of New York at Stony Brook, Stony Brook, NY 11794-3800, USA}
\affiliation{C. N. Yang Institute for Theoretical Physics, State University of New York at Stony Brook, Stony Brook, NY 11794-3840, USA}
\author{Shuyu Zhang}
\affiliation{Department of Physics and Astronomy, State University of New York at Stony Brook, Stony Brook, NY 11794-3800, USA}
\affiliation{C. N. Yang Institute for Theoretical Physics, State University of New York at Stony Brook, Stony Brook, NY 11794-3840, USA}
\author{Tzu-Chieh Wei}
\affiliation{Department of Physics and Astronomy, State University of New York at Stony Brook, Stony Brook, NY 11794-3800, USA}
\affiliation{C. N. Yang Institute for Theoretical Physics, State University of New York at Stony Brook, Stony Brook, NY 11794-3840, USA}
\begin{abstract}
    Quantum singular value transformation (QSVT) is a framework that has been shown to unify many primitives in quantum algorithms. In this work, we leverage the QSVT framework in two directions.
    We first \corr{revisit one recently introduced quantum power method within the QSVT framework and implement its matrix powers with block encodings, which makes the cost of this subroutine polylogarithmic in the inverse precision; the overall running time, however, remains exponential in the number of power iterations because of post-selection.} % [E7]
    Additionally, we incorporate several elementary numerical integration techniques, such as the rectangular method, Monte Carlo method, and quadrature method, into the QSVT framework, which results in 
    polynomial
    speedup with respect to the size or the number of points of the grid \corr{for the rectangular and quadrature rules, compared with exact classical summation over the grid}. Our results thus provide further examples to demonstrate the potential of the QSVT and how it may enhance quantum algorithmic tasks.
\end{abstract}
\maketitle
\section{Introduction}
Quantum computation has the capability to revolutionize the computational science frontier by incorporating the principle of quantum physics. Since the earliest attempt by Feynman~\cite{feynman2018simulating} and other pioneers, tremendous efforts have been devoted to exploring the power of a quantum computer in a wide array of areas~\cite{deutsch1985quantum, deutsch1992rapid, shor1999polynomial, regev2023efficient, berry2007efficient, berry2012black, berry2014high, berry2015hamiltonian, harrow2009quantum, lloyd2013quantum, mitarai2018quantum, lloyd2014quantum, lloyd2016quantum}. At the heart of quantum computation is the quantum algorithm. While many algorithms have been found in the above-cited references, a recent result~\cite{gilyen2019quantum}, the so-called quantum singular value transformation (QSVT), has revealed a unified perspective of quantum algorithms. \revise{This framework produces simpler and sometimes more cost-effective ways to execute certain primitives, such as the Hamiltonian simulation~\cite{low2017optimal, low2019hamiltonian, berry2007efficient,berry2012black,berry2014high, berry2015hamiltonian} and amplitude estimation~\cite{rall2020quantum, rall2023amplitude,rall2021faster}, eigenvalues finding, gradient descent~\cite{nghiem2023improved}, linear and nonlinear partial differential equations~\cite{krovi2023improved}, and so on.} Thus, it is desirable to explore further the potential of the QSVT framework and see to what extent the QSVT can be beneficial for what other tasks. 

Here, inspired by the aforementioned works, we make progress in two \revise{different} directions. First, building upon prior results in~\cite{gilyen2019quantum, nghiem2023quantum, nghiem2023improved}, we show \corr{how} the quantum power method proposed in~\cite{nghiem2023quantum} can be \corr{implemented with block encodings from the QSVT framework~\cite{gilyen2019quantum}}. \revise{Even though Ref.~\cite{nghiem2023improved} has already \corr{proposed a method} for finding the largest eigenvalue based on the original quantum power method~\cite{nghiem2023quantum},
the technique used there is more or less an implicit approach as the quantity of interest is not found directly but rather indirectly by solving a small linear system containing the variables, i.e., the eigenvalues of interest.} Here, we \corr{revisit} the result of~\cite{nghiem2023quantum} \corr{directly}, meanwhile keeping a similar routine, i.e., performing measurements to obtain a desired state, combined with the Hadamard test, except that \corr{the matrix powers are implemented with block encodings. The exponentially small probability of measuring the desired outcome, which was the main bottleneck in Ref.~\cite{nghiem2023quantum}, remains.} % [E6, E7] 
\revise{Then, switching to a different direction}, we attempt to develop elementary numerical integration techniques using the block encoding method. 
The key motivating idea is based on the recent results on preparing linear functions~\cite{gonzalez2024efficient}---by treating them as variables, and the QSVT allows us to transform these values with great flexibility. 
The numerical integration in such a domain can be estimated using the primitive amplitude estimation method. 
\revise{Though very different, the two topics serve as two important elementary mathematical operations used in many places, and their speedup can be potentially useful.}

The structure of the paper mainly consists of two parts. 
In Section~\ref{sec: acceleratedpower}, we begin by reviewing the original quantum power method introduced in~\cite{nghiem2023quantum}. We will explain the procedure and identify the main bottleneck of this approach. Then, we describe how \corr{block encodings from} the QSVT \corr{framework implement the corresponding task, and we derive its run time for a given number of power iterations}. % [E7] 
In the second part, to be described in Section~\ref{sec: integration}, we provide a method based on the QSVT that incorporates several well-known numerical integration schemes, such as the rectangular method, 
Monte Carlo method and quadrature method in the QSVT framework. 
\revise{As shown in previous works, such as~\cite{abrams1999fast} that used, e.g., a black box to compute functions, the quantum method indeed provides a certain speedup and memory saving to perform numerical integration. But in our work, we do not use such a black box.} Finally, we make some concluding remarks in Section~\ref{sec:conclude}.

\section{\texorpdfstring{\corr{Block-Encoding} Framework For Quantum Power Method}{Block-Encoding Framework For Quantum Power Method}} % [E7]
\label{sec: acceleratedpower}

We first review the original quantum power method developed in~\cite{nghiem2023quantum} to see how its bottleneck comes from the measurement and \corr{how block encodings from the QSVT framework can be used to implement it}. %
In this problem, the goal is to find the largest eigenvalue (in magnitude) of an $n\times n$ Hermitian matrix $A$ with sparsity $s$. 
We assume that the spectrum of $A$ is bounded \corr{in magnitude} between $1/\kappa$ (with $\kappa>\corr{2}$) and \corr{$1/2$} (in this case \corr{$\kappa/2$ bounds the condition number} of $A$);
otherwise, we can perform a trivial rescaling on the matrix. \corr{The upper bound $1/2$ provides the margin needed to remove the subnormalization of the block encoding in Lemma~\ref{lemma: newmatrixapp} below.} 

The power method begins with a random vector $x_0$, and then we apply $A$ to it at $k$ times. The resulting vector is denoted as $x_k = A^k x_0$. 
Let $\ket{x_k}$ be the normalized version of the vector $x_k$, i.e., $\ket{x_k} = x_k/|x_k|$, where $|.|$ refers to the familiar $l_2$ Euclidean norm. 
The largest eigenvalue $\lambda_{\rm max}$ of $A$ is estimated by the quantity $\bra{x_k} A \ket{x_k}$. The accuracy of the estimation clearly
depends on $k$. 

The matrix multiplication method essential for the quantum power method used in~\cite{nghiem2023quantum} is based on a data fitting method~\cite{wiebe2012quantum}, which is a modified Harrow-Hassidim-Lloyd (HHL) algorithm~\cite{harrow2009quantum}. For convenience, we quote the result as follows.
\begin{lemma}\label{lemma1}
  \revise{\bf (Matrix Multiplication)} Given oracle access to an $s$-sparse $n\times n$ Hermitian matrix $A$ and a random state $\ket{x_0}$. Suppose there is a unitary $U_{A^k}$ that performs the following:
   \begin{align}
        U  \ket{0}\ket{x_0} = \ket{0} A^k \ket{x_0} + \ket{1} \ket{\rm Garbage},
    \end{align}
    then there exists a quantum algorithm to realize a unitary $U$, i.e., a unitary $U_{A^k}$ performing: 
    \begin{align}
        U_{A^k} \ket{0}\ket{x_0} = \ket{0} \widetilde{A^k \ket{x_0} }+ \ket{1} \ket{\rm Garbage},
        \label{eqn: kiteration}
    \end{align}
with time complexity 
\begin{align*}
        \mathcal{O}\Big(  \frac{\log (n)s\kappa }{\epsilon} + \log(k)  \Big),
    \end{align*}
\revise{such that the achieved state $\widetilde{A^k \ket{x_0} }$ (unnormalized) well approximates the target state $A^k \ket{x_0}$ (unnormalized):  $|\widetilde{A^k \ket{x_0} } - A^k \ket{x_0} | \leq \epsilon$, where $|.|$ denotes the usual Euclidean norm of vectors, and $\ket{\rm Garbage}$ is an irrelevant state that is orthogonal to $A^k \ket{x_0}$}.
\end{lemma}
To obtain the desired state $\ket{x_k}$, one must measure the first register and post-select the outcome being $\ket{0}$. The probability of measuring $\ket{0}$ is $p_0 = |A^k \ket{x_0}|^2$ and, as shown in~\cite{nghiem2023quantum}, can be exponentially small in $k$, which is lower bounded by $(1/\kappa^{2k})$. 
Hence the running time of the method in~\cite{nghiem2023quantum} is
\revise{$ \mathcal{O}\Big(  \big(\frac{\log (n)s\kappa }{\epsilon} + \log(k)\big) \cdot \frac{\kappa^{2k} }{\delta^2} \Big)$, where $\delta$ is the additive error coming from measurements to the largest eigenvalue $\lambda_{\rm max}$. Note that the dependence of  $\kappa^{2k}$ 
in the run time can be improved to \corr{$\kappa^{k}$} by amplitude amplification to improve the scaling. We also note that the overall additive error to the eigenvalue then consists of both $\epsilon$ and $\delta$, and we can set $\epsilon={\cal O}(\epsilon_{\rm new})$ and $\delta={\cal O}(\epsilon_{\rm new})$, so there is only one error parameter in the final expression, which we will drop the subscript `new'.}  Thus, this is the bottleneck in the proposed algorithm. % [E23]

In~\cite{nghiem2023improved}, the authors proposed a way to overcome such an exponential scaling by using tools from~\cite{gilyen2019quantum}, such as Lemma~\ref{lemma: improveddme},  Lemma~\ref{lemma: product}, and Lemma~\ref{lemma: As} stated in Appendix~\ref{sec: prelim}. Instead of performing measurements directly to obtain the desired state, they estimate a few other variables that contain the quantity $\bra{x_k}A\ket{x_k}$ of interest, which is encoded in some linear combinations, and the solution reveals $\bra{x_k}A \ket{x_k}$. \corr{However, that construction is subject to the same issue as explained in the Remark below, so that the exponential scaling is not removed there either.}

Here, we \corr{follow a more direct approach} based on the QSVT framework. We \corr{use the following variant of} a lemma from~\cite{nghiem2023improved}.\footnote{
\revise{The quantum power method in Ref.~\cite{nghiem2023improved} focuses on finding the largest eigenvalue, and its \corr{stated} runtime complexity is $\mathcal{O}\big( k (s/\delta) (\log(n) + \log^{2.5} (1/\delta) )\big)$, where $\delta$ is the approximation error to the largest eigenvalue\corr{; this bound is subject to the issue explained in the Remark in Sec.~\ref{sec: acceleratedpower}}. The output of the algorithm of the present work also gives the eigenvector.}
} 
\begin{lemma}
\label{lemma: newmatrixapp}  \revise{{\bf (Matrix Multiplication)}}
Given oracle access to some $s$-sparse $n\times n$ Hermitian matrix $A$ as before, then the following unitary operation
\begin{align}
        U_{A^k} \ket{\bf 0}\ket{x_0} = \ket{\bf 0} \left(A\right)^k \ket{x_0} + \sum_{j \neq \textbf{0} } \ket{j}\ket{\rm Garbage}_j   
        \label{eqn: newmatrixapp}
\end{align}
can be realized up to additive accuracy $\epsilon$ (in a similar manner to lemma 1) in time 
$$ \corr{\mathcal{O}\left( s k \log\Big(\frac{sk}{\epsilon}\Big)\Big(\log(n) + \log^{2.5}\big(\frac{sk}{\epsilon}\big)  \Big)\right)}, $$ % [E4]
where $\ket{\bf 0}$ refers to an ancillary system.
\footnote{The ancillary system is of size $\mathcal{O}(k \log(n))$. In Lemma 48 of~\cite{gilyen2019quantum} (Lemma~\ref{lemma: As} in this paper), we need $\mathcal{O}(\log(n))$ ancillary qubits to construct the block encoding to $A/s$. In Lemma 53 of~\cite{gilyen2019quantum} (Lemma~\ref{lemma: product} in this paper), the number of ancillas adds up in each multiplication. In the rest of our construction for the quantum power method, the size of the ancillary system stays asymptotically the same.} 
We note that all $\{\ket{\rm Garbage}_j\}$ are not properly normalized.
\end{lemma}

\revise{We remark that both Lemma 1 and Lemma 2 are used to multiply some of a matrix to a vector. Lemma 1 was used in Ref.~\cite{nghiem2023quantum}, and Lemma 2 is \corr{a block-encoding version}~\cite{nghiem2023improved}, which we now derive in the block encoding framework  using Lemma~\ref{lemma: As}
to block encode matrix $A/s$,  remove the $s$ factor with Lemma~\ref{lemma: amp_amp}, and  Lemma~\ref{lemma: product} to block encode $A^k$ before apply it to $|x_0\rangle$.}
\begin{lemma}[\revise{\cite{gilyen2019quantum} Block Encoding of a  Matrix}]
\label{lemma: As}
    Given oracle access to $s$-sparse matrix $A$ of dimension $n\times n$, then an $\epsilon$-approximated unitary block encoding of $A/s$ can be prepared with gate/time complexity $\mathcal{O}\Big(\log n + \log^{2.5}(\frac{s^2}{\epsilon})\Big).$
\end{lemma}
This is presented in~\cite{gilyen2019quantum} (see their Lemma 48), and one can also find a review of the construction in~\cite{childs2017lecture}. We remark further that the scaling factor $s$ in the above lemma can be reduced by the preamplification method with further complexity \corr{$\mathcal{O}(s\log(s/\epsilon))$}~\cite{gilyen2019quantum}. % [E4]

\begin{lemma}[\cite{gilyen2019quantum} Theorem 30; \revise{\bf Amplification}]\label{lemma: amp_amp}
Let $U$, $\Pi$, $\widetilde{\Pi} \in {\rm End}(\mathcal{H}_U)$ be linear operators on $\mathcal{H}_U$ such that $U$ is a unitary, and $\Pi$, $\widetilde{\Pi}$ are orthogonal projectors. 
Let $\gamma>1$ and $\delta,\epsilon \in (0,\frac{1}{2})$. 
Suppose that $\widetilde{\Pi}U\Pi=W \Sigma V^\dagger=\sum_{i}\varsigma_i\ket{w_i}\bra{v_i}$ is a singular value decomposition. 
Then there is an $m= \mathcal{O} \Big(\frac{\gamma}{\delta}
\log \left(\frac{\gamma}{\epsilon} \right)\Big)$ and an efficiently computable $\Phi\in\mathbb{R}^m$ such that
\begin{equation}
\left(\bra{+}\otimes\widetilde{\Pi}_{\leq\frac{1-\delta}{\gamma}}\right)U_\Phi \left(\ket{+}\otimes\Pi_{\leq\frac{1-\delta}{\gamma}}\right)=\sum_{i\colon\varsigma_i\leq \frac{1-\delta}{\gamma} }\tilde{\varsigma}_i\ket{w_i}\bra{v_i} , \text{ where } \Big|\!\Big|\frac{\tilde{\varsigma}_i}{\gamma\varsigma_i}-1 \Big|\!\Big|\leq \epsilon.
\end{equation}
Moreover, $U_\Phi$ can be implemented using a single ancilla qubit with $m$ uses of $U$ and $U^\dagger$, $m$ uses of C$_\Pi$NOT and $m$ uses of C$_{\widetilde{\Pi}}$NOT gates and $m$ single qubit gates.
Here,
\begin{itemize}
\item C$_\Pi$NOT$:=X \otimes \Pi + I \otimes (I - \Pi)$ and a similar definition for C$_{\widetilde{\Pi}}$NOT; see Definition 2 in \cite{gilyen2019quantum},
\item $U_\Phi$: alternating phase modulation sequence; see Definition 15 in \cite{gilyen2019quantum},
\item $\Pi_{\leq \delta}$, $\widetilde{\Pi}_{\leq \delta}$: singular value threshold projectors; see Definition 24 in \cite{gilyen2019quantum}.
\end{itemize}
\end{lemma}

\revise{
\begin{lemma}[Block Encoding of Product of Two Matrices]
\label{lemma: product}
    Given the unitary block encoding of two matrices $A_1$ and $A_2$, then there exists an efficient procedure that constructs a unitary block encoding of $A_1 A_2$ using each block encoding of $A_1,A_2$ one time. 
\end{lemma}
The proof of the above lemma is given in \cite{gilyen2019quantum}, see their Lemma 53. }

\medskip 

\corr{To reproduce Lemma~\ref{lemma: newmatrixapp} using block encoding, the first step is to construct an $(\epsilon/s)$-close block encoding of $A/s$ in time $\mathcal{O}(\log(n) + \log^{2.5} (s/\epsilon))$ by Lemma~\ref{lemma: As}. Since the singular values of the encoded block are at most $1/(2s)+\epsilon/s\le 3/(4s)$ (for $\epsilon\le 1/4$), Lemma~\ref{lemma: amp_amp} with $\gamma=s$ and $\delta=1/4$ applies and removes the factor $s$: it yields an $\epsilon$-close block encoding $U_{A}$ of $A$ (after rescaling $\epsilon$ by a constant factor), incurring further complexity $\mathcal{O}(s\log(s/\epsilon))$ (for $s=1$, no amplification is needed).} % [E4]
Then, Lemma~\ref{lemma: product} allows for a recursive construction of $U_{A^{\ell +1}}$, which is an $(\ell+1)\epsilon$-close block encoding of $(A)^{\ell+1}$, by multiplication $(I_a \otimes U_{A}) (I_{a'} \otimes U_{A^\ell})$, where $U_{A^{\ell}}$ is an $\ell \epsilon$-close block encoding of $(A)^{\ell}$ and ancillas $a$ and $a'$ supporting the identity matrices are set appropriately.
The $k$ multiplications of the encoding unitary matrix will increase the error linearly in $k$, so we need to rescale it as $k\epsilon \rightarrow \epsilon$, yielding the $\frac{k}{\epsilon}$ factor in the logarithm.

\begin{corrblock}
\medskip\noindent\textit{Remark (correction to the published version).} In the published version of this paper [Phys.\ Rev.\ A \textbf{111}, 012434 (2025)], we block encoded the rank-one operator $\widetilde{\rho}_k=\zeta_k\ket{x_k}\bra{x_0}$ with $\zeta_k=(|x_k|^2/s^{2k})\braket{x_k|x_0}$, whose only nonzero singular value $|\zeta_k|$ is exponentially small in $k$, and applied a polynomial approximation of $\frac12 e^{-\beta(1-x)}$ by means of the polynomial transformation of Lemma~\ref{lemma: theorem56} in order to boost the success probability of the post-selection to a constant. This step is \emph{incorrect}. Lemma~\ref{lemma: theorem56} (Theorem 56 of~\cite{gilyen2019quantum}) transforms the eigenvalues of a Hermitian matrix, whereas $\widetilde{\rho}_k$ is not Hermitian. For a non-Hermitian operator, the singular values can be transformed only by polynomials of definite parity. Splitting the polynomial into its odd and even parts, the odd part, $\frac12 e^{-\beta}\sinh(\beta x)$, maps $\ket{x_0}$ to $\frac12 e^{-\beta}\sinh(\beta|\zeta_k|)\ket{x_k}$, whose amplitude remains exponentially small, while the even part, $\frac12 e^{-\beta}\cosh(\beta x)$, acts on the right singular vectors and maps $\ket{x_0}$ to approximately $\frac12\ket{x_0}$; the constant success probability obtained there thus corresponds to the state $\ket{x_0}$ rather than $\ket{x_k}$. More generally, an odd polynomial $P$ with $|P(x)|\le1$ on $[-1,1]$ satisfies $|P(\sigma)|\le \deg(P)\arcsin(\sigma)$ by Bernstein's inequality, so that boosting a singular value $\sigma$ to a constant requires $\deg(P)=\Omega(1/\sigma)$. Consequently, the claims of the published version that the success probability is boosted to a constant and that the runtime is polylogarithmic in $n$ and independent of $\kappa$ (Theorem 1 and Table I there) are incorrect. Similar errors appeared in Refs.~\cite{nghiem2023improved,nghiem2025improved}, and the incorrect complexity of largest-eigenvalue finding was used in Refs.~\cite{nghiem2025quantumAlgo,nghiem2026quantum}.  
Here, we gratefully acknowledge Jens Renders for pointing out a mistake in our previous version. 
The corrected approach below avoids the singular value transformation of $\widetilde{\rho}_k$ and applies the block-encoded matrix power directly, as in Ref.~\cite{nghiem2023quantum}.
\end{corrblock}

\begin{corrblock}

\smallskip
\noindent\textit{Corrected procedure.} By applying the block-encoded $A^k$ of Lemma~\ref{lemma: newmatrixapp} to the state $\ket{\bf 0}\ket{x_0}$, where $\ket{x_0}$ can be prepared, e.g., as a random product state with $\mathcal{O}(\log n)$ gates, we obtain the state
\begin{align}
    \ket{\bf 0}\, \widetilde{A^k}\ket{x_0} + \sum_{j \neq \textbf{0} } \ket{j}\ket{\rm Garbage}_j,
    \label{eq: corrected-state}
\end{align}
where $|\widetilde{A^k}\ket{x_0} - A^k\ket{x_0}|\le\epsilon'$ for the accuracy $\epsilon'$ chosen in Lemma~\ref{lemma: newmatrixapp}. Measuring the ancilla and post-selecting on $\ket{\bf 0}$ yields the state $\ket{x_k}$ up to the error discussed below, which is an approximation to the eigenvector (with largest eigenvalue) of $A$. Writing $\ket{x_0}=\sum_i c_i\ket{E_i}$ in the eigenbasis $\{\lambda_i,\ket{E_i}\}$ of $A$, the success amplitude in the ideal case is $|x_k| = (\sum_i |\lambda_i|^{2k}|c_i|^2)^{1/2}\ge\kappa^{-k}$, which is exponentially small in $k$, as in Ref.~\cite{nghiem2023quantum}. Because the post-selection normalizes the state by this amplitude, the error $\epsilon'$ is amplified to at most $2\epsilon'\kappa^{k}$ in the post-selected state; we therefore choose $\epsilon'=\epsilon\kappa^{-k}/4$, which enters the runtime only logarithmically. Instead of repeating the measurement, we use fixed-point amplitude amplification (Theorem 27 of~\cite{gilyen2019quantum}) with the lower bound $\kappa^{-k}/2$ on the success amplitude, which prepares an $\epsilon$-approximation of $\ket{x_k}$ with $\mathcal{O}(\kappa^{k}\log(1/\epsilon))$ uses of $U_{A^k}$, $U_{A^k}^\dagger$, and the unitary preparing $\ket{x_0}$. With $\ket{x_k}$, we follow the same routine as outlined originally in~\cite{nghiem2023quantum} and execute the Hadamard test, with the controlled version of the block encoding $U_A$ of $A$ constructed above (built by taking the controlled versions of all the gates within this unitary), combined with amplitude estimation; this reveals $\bra{x_k}A\ket{x_k}$ to additive error $\epsilon$ with $\mathcal{O}(1/\epsilon)$ repetitions. The errors from the block encodings, the amplitude amplification, and the Hadamard test add up to $\mathcal{O}(\epsilon)$, and we rescale $\epsilon$ by a constant factor.
\end{corrblock}

We summarize our result in the following theorem.
\begin{theorem}[\corr{Quantum Power Method with Block Encodings}]
\label{thm: main}
 Given oracle access to an s-sparse Hermitian matrix $A$ \corr{whose eigenvalues satisfy $1/\kappa\le|\lambda_i|\le 1/2$, the quantum power method with any $k$ iterations, which prepares the state $\ket{x_k}$ and estimates $\bra{x_k}A\ket{x_k}$ to additive error $\epsilon$ with constant success probability,} can be executed in \revise{run time} complexity
 \begin{corrblock}
 $$ \mathcal{O}\Big( \kappa^{k}\, \frac{s}{\epsilon}\, k\, \Big[\log(n) + \log^{2.5}(L)\Big]\log(L)\, \log\Big(\frac{1}{\epsilon}\Big)\Big), \qquad L=\frac{sk\kappa^{k}}{\epsilon}. $$
 \end{corrblock}
\end{theorem}
% [E3]

\corr{The factor $\kappa^{k}$ stems from the post-selection: its success probability $|x_k|^2\ge\kappa^{-2k}$ is exponentially small in $k$, and amplitude amplification reduces its cost only quadratically, as in Ref.~\cite{nghiem2023quantum}. Hence, the runtime bound is exponential in the number of iterations $k$. Compared with Lemma~\ref{lemma1} used in Ref.~\cite{nghiem2023quantum}, the block-encoding implementation of the matrix power (Lemma~\ref{lemma: newmatrixapp}) has a cost polylogarithmic, rather than linear, in $1/\epsilon$.} % [E5]

\section{QSVT-based Framework For Numerical Integration}
\label{sec: integration}

Here, we switch to a different subject that we will also improve using the QSVT framework. Numerical integration finds numerous applications in many areas. Standard textbooks such as~\cite{hamming2012numerical} \revise{contain  thorough details on various methods and their error analysis}. In this section, we explore the integration of \revise{single- and} multi-variable polynomials over certain domains. 
Such integration is elementary and, indeed, can be performed analytically. 
The aim of this section is to explore the capability of quantum computers in this simple setup.

\revise{Most numerical integration methods break the interval (or domains) into multiple parts, e.g., a uniformly spaced grid, and the desired function, as well as integration, is approximated accordingly on the grid}. The accuracy,  usually quantified by the $l_2$-norm deviation, depends on the number of subintervals (or subdomains) presented. 
Typically, the more, the better, but it increases the computational cost in both running time and hardware resources. 
The topic of numerical integration has \revise{previously} been addressed from the quantum perspective in several works~\cite{abrams1999fast, novak2001quantum,shu2024general}, suggesting that quantum computers can provide a potentially useful framework, for example, to represent a function at $n$ points on the given interval, a quantum computer requires $\log(n)$ qubits only. 
\revise{Here, we employ the QSVT framework, which 
 enables one to deal with polynomial functions in a natural and systematic way.} Meanwhile, it is well-known from the context of analysis that any continuous function can be approximated by polynomials within an appropriate domain. 
In~\cite{abrams1999fast}, the authors consider the black box model that computes the unknown function upon some queries. In our work here, instead of the black box model, we consider an explicit function, as we are inspired by the following features. First, in some closed interval, any continuous function can be approximated by certain polynomials. Second, the QSVT allows us to perform such a polynomial transformation efficiently. In fact, these two features were also exploited in the work of~\cite{gonzalez2024efficient}, where the authors consider the problem of loading functions into a quantum state. Our work goes further by integrating the desired multivariate functions over some regions. In the following, for simplicity, we mainly choose the region of interest to be $\mathscr{D} = [-0.5,0.5]^d$, where $d \in \mathbb{Z}_+$ is the spatial dimension; however, it is not essential.

\subsection{Integration of Polynomial Function}
\begin{figure}
\includegraphics[width=0.7\linewidth]{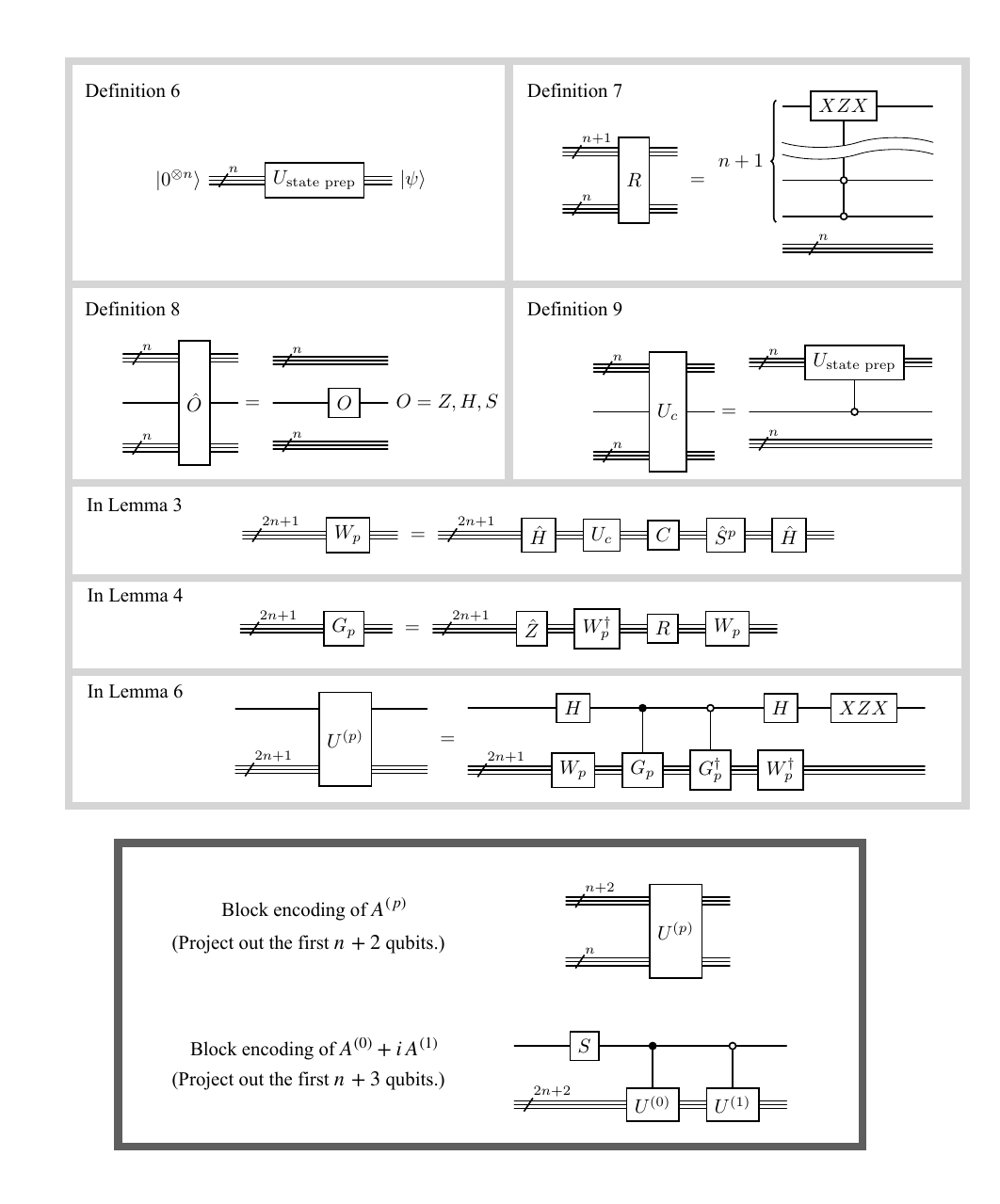}
\caption{\revise{Review of the quantum circuit in Ref.~\cite{rattew2023non} used in Lemma~\ref{lemma: diagonal}.
The top table lists definitions according to the numbering of Ref.~\cite{rattew2023non}. 
Here, the gate $C$ is a controlled-copy gate that transfers the last $n$ entries into the first $n$ entries conditioned by the middle single qubit: $C = 
\sum_k |0^{\otimes n}\rangle \langle 0^{\otimes n}| \otimes | 0 \rangle \langle 0| \otimes |k\rangle_n \langle k |_n + |k\rangle \langle 0^{\otimes n}| \otimes | 1 \rangle \langle 1| \otimes |k\rangle_n \langle k |_n$.
The block encoding of $A$ is achieved separately for its real part $A^{(0)}$ and imaginary part $A^{(1)}$. 
They are block encoded into $U^{(0)}$ and $U^{(1)}$, respectively. Each of them consists of gadgets $W_p$ and $G_p$ which are build upon the state preparation unitary $U_\text{state prep}$ that loads the wave function $|\psi\rangle$.}}
\label{fig:from_wave_function_to_diagonal}
\end{figure}

We first construct a recipe useful for subsequent discussions on numerical integration. First, we call a result from~\cite{rattew2023non}, which allows us to block encode a diagonal matrix from a vector:
\begin{lemma}[Theorem 2 in \cite{rattew2023non}; \revise{Block Encoding into a Diagonal Matrix}]
\label{lemma: diagonal}
     Given an n-qubit quantum state specified by a state-preparation-unitary $U$, such that $\ket{\psi}_n=U\ket{0}_n=\sum^{N-1}_{k=0}\psi_k \ket{k}_n$ (with $\psi_k \in \mathbb{C}$ and $N=2^n$), we can prepare an exact block-encoding $U_A$ of the diagonal matrix $A = {\rm diag}(\psi_0, ...,\psi_{N-1})$ with $\mathcal{O}(n)$ circuit depth and a total of $\mathcal{O}(1)$ queries to a controlled-$U$ gate  with $n+3$ ancillary qubits.
\end{lemma}
\revise{We give an illustration of quantum circuits used in this Lemma in Fig.~\ref{fig:from_wave_function_to_diagonal}.}
The above tool can be applied directly to obtain the following:
\begin{lemma}
\label{lemma: uniformloading} \revise{\bf (Block Encoding of a Linear Function into a Diagonal Matrix)}
    There exists a quantum circuit of depth $\mathcal{O}(N^{1/2}\log(N/\epsilon) \log(N) )$ that generates an $\epsilon$-close block encoding of the following matrix \footnote{\revise{The factor $-N/2$ appears to enforce the requirement that the range of discrete variables is within $(-1/2,1/2)$. We could also set it to be $(0,1)$ instead; particular choice of the integration range is not important so long as it is bounded by some constant.}}:
    \begin{equation}
    A = \frac{1}{N} \sum_{j=0}^{N-1} \Big(j-\frac{N}{2} \Big) \ket{j}\bra{j}.
    \label{A}
    \end{equation}
\end{lemma}
\textit{Proof:} First let $n = \log_2(N)$ (in this lemma only) for brevity. We have that $H^{\otimes n} \ket{0}^n = (1/\sqrt{N}) \sum_{j=0}^{N-1} \ket{j}$. We then use an extra ancilla $\ket{0}$ and perform the following rotation denoted by $U_R$, 
\begin{align}
\label{eqn: rotation}
    \ket{0} \ket{j} \longrightarrow  \Bigg( \frac{2j-N}{2N} \ket{0} + \sqrt{1 - \frac{(2j-N)^2}{4N^2}} \ket{1}   \Bigg) \ket{j},
\end{align}
which has complexity $\mathcal{O}(\log(N))$ \revise{(see Ref.~\cite{schuld2018supervised} Section 5.2.2 Qubit-Efficient State Preparation and Ref.~\cite{grover2000synthesis})}\corr{; more precisely, since the rotation angle is not linear in $j$, it is computed with reversible arithmetic to $\mathcal{O}(\log(N/\epsilon))$ bits~\cite{grover2000synthesis}, which costs $\mathrm{polylog}(N/\epsilon)$. For brevity, we keep writing $\mathcal{O}(\log(N))$ for this cost below, which affects only polylogarithmic factors}. We then have the following (assuming $N=2^n$):
\begin{align}\label{eq:rotation}
I\otimes H^{\otimes n}|0\rangle \otimes |0\rangle^{n}   = \frac{1}{\sqrt{N}} \sum_{j=0}^{N-1} \ket{0} \ket{j} 
\xrightarrow{U_R}
 \frac{1}{\sqrt{N}} \sum_{j=0}^{N-1} \Bigg( \frac{2j-N}{2N} \ket{0}\ket{j} + \sqrt{1 - \frac{(2j-N)^2}{4N^2}} \ket{1}\ket{j} \Bigg) .
\end{align}
Therefore, if we use Lemma~\ref{lemma: diagonal} to the r.h.s. of the above, i.e., setting $U=U_R\cdot I\otimes H^{\otimes n} $ there, then we obtain a block encoding of the following diagonal matrix 
\begin{equation}
\sum_{j=0}^{N-1} \frac{1}{\sqrt{N}} \frac{2j-N}{2N} \ket{0}\bra{0} \otimes \ket{j}\bra{j} + \sum_{j=0}^{N-1} \Bigg( \frac{1}{\sqrt{N}}\sqrt{1-\frac{(2j-N)^2}{4N^2}} \Bigg) \ket{1}\bra{1} \otimes \ket{j}\bra{j} 
\end{equation}
with a circuit of depth $\mathcal{O}(\log(N))$. 
It is simple to see that the first part is a block encoding of $\sum_{j=0}^{N-1} (2j-N)/(2N\sqrt{N}) \ket{j}\bra{j}$, which means that we already obtain a block encoding of $A/\sqrt{N}$, where $A$ is defined as in equation~(\ref{A}). 
In order to remove the factor $\sqrt{N}$, one can use the amplification
technique in Ref.~\cite{gilyen2019quantum} (Lemma~\ref{lemma: amp_amp}).

We choose the amplification factor $\gamma = \sqrt{N}$ and the truncation width $\delta = \corr{\frac{1}{4}}$. Note this restricts eigenvalues \corr{to $|(2j-N)/2N\sqrt{N}|\le 3/(4\sqrt{N})$}, which is automatically satisfied in our case\corr{, since $|(2j-N)/2N\sqrt{N}|\le 1/(2\sqrt{N})$}. % [E20]
Such a unitary with amplified singular values can be implemented with $m=\mathcal{O}(\frac{\gamma}{\delta} \log\big(\frac{\gamma}{\epsilon}\big))$ uses of the pre-amplified unitaries (i.e., $U_{\rm diag}$) and other gates. 
This completes Lemma~\ref{lemma: uniformloading}.

Subsequently, we will build from these results to deal with multi-variable settings. Here, we briefly point out a similar result in~\cite{gonzalez2024efficient}, where the authors show how to load a linear function. 
\begin{lemma}[\cite{gonzalez2024efficient} \revise{Loading of a Linear Function into a Diagonal Matrix}]
\label{lemma: linearloading}
    There exists a quantum circuit of depth $\mathcal{O}( \log(N) )$ that generates an exact block encoding of $A = \frac{1}{C}\sum_{j=0}^{N-1} j \ket{j}\bra{j}$, where $C = \sqrt{ (2N-1) \ ( N - 1) \ N /6  }$. 
\end{lemma}

\revise{
Note that, in Lemma~\ref{lemma: uniformloading} and Lemma~\ref{lemma: linearloading}, scaling factors differ as $C$ versus $N$.
In our one-dimensional integration, we restrict to the range $(-1/2,1/2)$, the linear function over which is achieved in Lemma~\ref{lemma: uniformloading}. 
On contrary, Lemma~\ref{lemma: linearloading} gives a linear function from $0$ to $\mathcal{O}(1/\sqrt{N})$, and adjusting it to the range $(-1/2,1/2)$ requires additional amplification by the scaling factor $\mathcal{O}(\sqrt{N})$, etc.
Hence, the latter approach achieves the linear function over the same range with the same complexity.
} 
The following tool is another one to be used in our subsequent construction, which is basically the polynomial transformation from the seminal work~\cite{gilyen2019quantum}, see also Theorem 27.2 in~\cite{childs2017lecture}.
\begin{lemma}[\cite{gilyen2019quantum} Theorem 56; Polynomial Transformation]
\label{lemma: theorem56}\label{lemma: qsvt}
Suppose that $U$ is an $(\alpha,a,\epsilon)$ encoding of a Hermitian matrix $A$ (see Definition 43 of~\cite{gilyen2019quantum} for the definition). If $P\in\mathbb{R}[x]$ is a degree-$d$ polynomial satisfying $|P(x)|\le\frac12$ for all $x\in[-1,1]$, then there is a quantum circuit $\widetilde{U}$ that is a $(1,a+2,4d\sqrt{\epsilon/\alpha})$ encoding of $P(A/\alpha)$ and consists of $d$ applications of $U$ and $U^\dagger$ gates, a single application of the controlled-$U$ gate, and $\mathcal{O}((a+1)d)$ applications of other one- and two-qubit gates. \corr{If, in addition, $P$ has a definite parity, the condition can be relaxed to $|P(x)|\le1$ for all $x\in[-1,1]$, with the same error bound (Corollary 18 of~\cite{gilyen2019quantum}).}
\end{lemma}

As mentioned in~\cite{rattew2023non}, generalizing the above lemma to a multi-variable setting is straightforward, but it lacks a detailed procedure there. Rather than constructing a single block encoding of matrices containing products of the multi-variables, we construct block encodings of each variable individually, and block encoding arithmetic allows us to build more complex multivariate functions. A concrete example is provided in appendix \ref{sec: example}. We summarize the desired tool in the lemma below.

\revise{To prepare for the lemma, let us consider a $d$-variable polynomial $f\in \mathbb{R}[x^{(1)},\cdots, x^{(d)}]$ (properly normalized\corr{, i.e., $f=\sum_{\vec p}C_{\vec p}\,(x^{(1)})^{p_1}\cdots(x^{(d)})^{p_d}$ with $\sum_{\vec p}|C_{\vec p}|\le1$}). % [E19]
We consider sampling some points $\Vec{x}_j=(x^{(1)}_j,\cdots , x^{(d)}_j)_{j=1,\cdots, N}\in \mathbb{R}^{d}$.
We provide a QSVT construction of such sampling in a polynomial as follows:}
\begin{lemma}[Loading of Multi-Variate Polynomial Function On Grid]
\label{lemma: gridpoly}
  Given the $\epsilon$-approximation block encoding of multiple diagonal matrices $X^{\revise{(i)}} = \sum^{\revise{N}}_{j=0}
  x^{(i)}_j \ket{j}\bra{j}$ for various $i = 1,2,.., d$ representing the discretized variables on grid, such as $(2j-N)/2N$ for $j=0,1,...,N-1$ for 1-d grid case restricted on domain $(-0.5,0.5)$, under consideration and a multi-variate polynomial $f: \mathbb{R}^d \rightarrow \mathbb{R}$ of interest. There exists a procedure with time complexity 
  \revise{$\mathcal{O}(N^{1/2}\log(N d {\rm deg}(f)  /\epsilon) \log(N) {\rm deg} (f) t(f))$} 
  (where $\deg(f)$ is the degree of polynomial $f$ and $t(f)$ is the number of terms in the polynomial $f$) that constructs a block encoding of a diagonal matrix that contains $f(x^{(1)}_{i_1},x^{(2)}_{j_2},...,x^{(d)}_{k_d})$ for all values of different variables 
  \revise{
  $\{x^{(i)}_j\}\vert_{j=1}^N$} and $i=1,2,...,d$.
  The number of ancillary qubits required is   
\revise{$\mathcal{O}(\log(N)d + \log(t(f)))$.}
\end{lemma}
The proof of the Lemma is presented in Appendix~\ref{sec: example}, together with an example.

\medskip\noindent \revise{{\bf Overview of the integration method}. The general procedure of numerical integration within quantum singular value transformation framework (1) starts with Lemma~\ref{lemma: uniformloading} or Lemma~\ref{lemma: linearloading} to load the arguments into a diagonal block-encoded matrix and (2) then proceeds with Lemma~\ref{lemma: gridpoly} above by using the QSVT to transform each argument into its functional value 
$f(x^{(1)}_{i_1},x^{(2)}_{j_2},...,x^{(d)}_{k_d})$. (3) Finally, measurements are performed to read out the summation of these functional values, i.e., using amplitude estimation.
Now, we are ready to show some concrete examples with well-known numerical integration schemes from the block encoding of polynomials that we have just described.}

\subsubsection{Integration According to Rectangular Rule}
In this approximation scheme, the domain $\mathscr{D}$ is broken into grids of equal size, and the given function is approximated at each point in the grid. The integration of some real-valued function $f: \mathbb{R}^d \rightarrow \mathbb{R}$ over some domain $\mathscr{D}$ is approximated by:
\begin{align}
    \int_\mathscr{D} f(\xbf) d\xbf = \frac{ \prod_{j=1}^d L_i }{M} \sum_i f(\xbf_i), 
\end{align}
where $L_i$ is the length of the given interval in the $i$-th dimension and $M$ is the number of points in the grid. As mentioned earlier, we choose $\mathscr{D} = (-1/2,1/2)^d$ for simplicity and thus the volume is unity. Furthermore, from Lemma~\ref{lemma: uniformloading}, we naturally obtain a grid (in 1-d) having $N$ points. In $d$ dimensions, we have $M = N^d$ points. 

As mentioned, over some closed interval (which can be of high dimensionality), a continuous function can be approximated by some polynomials, and if the domain $\mathscr{D}$ is too large, in practice, we can break $\mathscr{D}$ into many smaller domains $\mathscr{D} = \bigcup_{i} \mathscr{D}_i $, integrate within each domain $\mathscr{D}_i$ and sum the results over. Lemma~\ref{lemma: gridpoly} allows us to load the values of the desired multi-variate polynomial into the diagonal entries of some matrix $P(A)$, which is block encoded by a unitary, denoted as $U_A$. In order to estimate the integration, we note that:
\begin{align}
    \bra{0}^{ \otimes \log(M)} H^{\otimes \log(M)} \ P(A) \ H^{\otimes \log(M)} \ket{0}^{\otimes \log(M)} = \frac{1}{M} \sum_{i=1}^M \bra{i} \ P(A) \ \ket{i} 
    = \frac{1}{M} \sum_i P(\xbf_i),
    \label{eq: sandwich}
\end{align}
which is exactly the desired integral that we want to perform. It is useful to note that the above numerical integral is exactly the amplitude, e.g., square root probability of measuring zeros in the computational basis state, of the state $\mathbb{I} \otimes H^{\otimes \log(M) } \cdot U_A \cdot \mathbb{I} \otimes H^{\otimes \log(M)} \ket{\bf 0}\ket{0}^{\otimes \log(M)}$. To see this, we observe that from the definition of block encoding~(\ref{def: blockencode}) (and its property in equation~\ref{eqn: action}):
\begin{align}
     \ket{\Phi} &\equiv\big(\mathbb{I} \otimes H^{\otimes \log(M)}\big)   U_A \big( \mathbb{I} \otimes H^{\otimes \log(M)}\big) \ket{\bf 0} \ket{0}^{\otimes \log(M)} \\ 
  &= \mathbb{I} \otimes H^{\otimes \log(M)} U_A \ket{\bf 0} H^{\otimes \log(M)} \ket{0}^{\otimes \log(M)} \\
    &= (\mathbb{I} \otimes H^{\otimes \log(M)})  (\ket{\bf 0} P(A) H^{\otimes \log(M)} \ket{0}^{\otimes \log(M)}  + \ket{\rm Garbage} ) \\
    &= \ket{\bf 0} H^{\otimes \log(M)} P(A) H^{\otimes \log(M)}\ket{0}^{\otimes \log(M)} + (\mathbb{I} \otimes H^{\otimes \log(M)}) \ket{\rm Garbage} .
\end{align}
It is clear from the above expression that the amplitude $\bra{\bf 0}\bra{0}^{\otimes \log(M)} \ket{\Phi}$ is our integral of interest, e.g., $\bra{0}^{ \otimes \log(M)} H^{\otimes \log(M)} \\ P(A) \ H^{\otimes \log(M)} \ket{0}^{\otimes \log(M)} $, which can be estimated using amplitude estimation. \\

\noindent\textit{\textbf{Comparing with classical method:}}  Classically, one needs to perform the summation over all values defined on the grid in dimension $d$, which has $M = N^d$ points. 
Therefore, the classical complexity is $\mathcal{O}(\deg(f) N^d )$ where $\deg(f)$ is the degree of the polynomial under consideration. 
In the quantum case, the complexity  is
\revise{$\mathcal{O}\Big(N^{1/2}\log\big(N d\, {\rm deg}(f)  /\epsilon\big) \log(N) {\rm deg} (f) t(f)/\epsilon\Big)$} 
where $t(f)$ is the number of terms in $f$, mainly due to the implementation of the degree-$d$ polynomial (as the gate $H^{\log(M)}$ can be implemented with a single layer), as well as the construction of the block encoding of discrete variables,  and $\epsilon$ is the error tolerance coming from amplitude estimation. The speedup is polynomial with respect to $N$\corr{, compared with exact classical summation. We note that the quantum algorithm outputs an $\epsilon$-additive estimate; for this task, a classical randomized algorithm that averages $f$ (with $|f|\le1$) over $\mathcal{O}(1/\epsilon^2)$ uniformly random grid points has a cost independent of $N$, so that, compared with it, the quantum advantage is at most quadratic in $1/\epsilon$, as in quantum versions of Monte Carlo methods~\cite{montanaro2015quantum}
}. In addition, the quantum \corr{framework} requires
\revise{$\mathcal{O}(\log(N)d + \log(t(f)))$}
qubits to store and process the procedure; \corr{we note, however, that the classical summation can also be carried out in a ``streaming'' fashion, without storing the grid points, using $\mathcal{O}(d\log N)$ bits for the grid index and a constant number of words for the accumulator, so that there is no exponential memory saving.} 

\subsubsection{Integration According to Monte Carlo Method}
\noindent \textit{\textbf{Uniform Sampling:}} The above rectangular rule is a definite approach where the value of a function is evaluated at specific points, e.g., on the uniform grid. Another popular method is the Monte Carlo method, which is probabilistic, and its quantum version has appeared in Ref.~\cite{montanaro2015quantum}. The method begins with a set of random points chosen from the domain $\mathscr{D}$ according to a uniform probability distribution. The function $f$ is then evaluated at and averaged over those points. The integration of $f$ over such a domain is then given by:
\begin{align}
    \int_\mathscr{D} f(\xbf) d\xbf = \frac{|\mathscr{D}|}{M} \sum_{i=1}^M f(\xbf_i),
\end{align}
where $M$ is the number of points and $|\mathscr{D}|$ is the volume of domain $\mathscr{D}$. The practical efficacy of this method depends a lot on the local behavior of the function, the number of chosen points, and as well as the number of repetitions (if necessary). At first glance, it seems that the Monte Carlo integration can be executed in a very natural manner in the quantum context, as it is probabilistic in essence. However, the generation of a random state does not really have its amplitude obeying a uniform distribution. To incorporate the Monte Carlo technique into the quantum setting, we propose the following procedure. 
 
For simplicity, we begin with the 1D case. Higher dimensions can be generalized straightforwardly, as we just need to randomize points (or coordinates) in all $d$ dimensions and use the same technique as outlined before Lemma~\ref{lemma: gridpoly} to deal with multi-variate settings. First, we classically randomize $M$ points, with coordinates as $\{x_1',x_2',...,x_M'\}$ between $(-1/2,1/2)$. We then classically perform the normalization for all points:
\begin{align*}
     x_i' \longrightarrow x_i = \frac{x_i'}{\sqrt{ \sum_{i=1}^M x'^2_i }}. 
\end{align*}
Then, we employ the amplitude encoding method, for example, as outlined in Section~2.2 of~\cite{prakash2014quantum}, which is basically summarized as follows. \corr{Given a quantum random access memory (QRAM) that stores the $b$ known amplitudes in a binary-tree data structure~\cite{prakash2014quantum}, a quantum state in $\mathbb{R}^b$ can be prepared in time $\mathcal{O}(\mathrm{polylog}(b))$; without such a QRAM, preparing a generic state with $b$ given amplitudes requires $\Omega(b)$ gates.
However, with use of exponentially many ancillas, a quantum state with known amplitudes can be efficiently prepared~\cite{zhang2022quantum,zhang2021lowdepthquantumstatepreparation}. But we will not consider such excessive use of ancillas here.}  
Therefore, the following state % [E11]
\begin{align}
    \ket{\phi} = \sum_{i=1}^M x_i \ket{i}
\end{align}
can be prepared \corr{in time $\mathcal{O}(\mathrm{polylog}(M))$, given a QRAM storing the $M$ points}. Using Lemma~\ref{lemma: diagonal}, one can prepare the block encoding of the operator:
$$  A = \sum_{i=1}^M x_i  \ket{i}\bra{i}, $$
where the diagonal entries of $A$ are random points $\{x_i'\}$  up to a normalization factor $\sqrt{\sum_{i=1}^M x_i'^2}$.
One can use the amplification to remove such a factor with a further complexity $\mathcal{O}( \sqrt{\sum_{i=1}^M x_i'^2})$. Then, one can follow the same procedure as the above grid-based method, e.g., use the method in Lemma~\ref{lemma: theorem56} to transform the diagonal entries and perform the Hadamard test to reveal the integration. 
Due to the use of amplification to remove the normalization factor, there is an additional complexity $\mathcal{O}(\sqrt{M}\log(M/\epsilon))$. \revise{In $d$ dimensions, we need to simply incorporate more variables, and lemma \ref{lemma: gridpoly} can be used in a straightforward manner to construct a block encoding of multivariate functions}. Thus, the total complexity is 
\revise{$\mathcal{O}\Big( \sqrt{M} \log\big(M d\, {\rm{deg}}(f)/\epsilon\big) \log(M) \deg(f) t(f)/\epsilon \Big)$}, similar to the result in the previous section \corr{(given the QRAM)}. \\

\noindent \textit{\textbf{Importance Sampling:}} The above uniform sampling draws points with equal probability. Another approach for Monte Carlo integration is the importance sampling, which draws points according to the chosen probability distribution. Denote the probability density as $g(x)$, and suppose we have $M$ points $\{x_1',x_2',...,x_M'\}$ chosen according to such density function (still we are working in 1-d, for simplicity). Then we have that the integral can be estimated numerically as follows,:
\begin{align}
    \int_\mathscr{D} f(x) dx = \frac{1}{M} \sum_{i=1}^M \frac{f(x_i')}{g(x_i')}.
\end{align}
In order to proceed, we first rescale the variable to the range $(-1/2,1/2)$ by performing the normalization as the previous steps, i.e., $x_i' \longrightarrow x_i = x_i'/ \sqrt{\sum_{i=1}^M x_i'^2 }$. We then construct the same matrix:
$$  A = \sum_{i=1}^M x_i \ket{i}\bra{i} = \sum_{i=1}^M \frac{x_i'}{ \sqrt{\sum_{i=1}^M x_i'^2 }} \ket{i}\bra{i},$$
and use amplification to remove the factor $\sqrt{\sum_{i=1}^M x_i'^2}$. \corr{Since $|x_i'|\le1/2$, we have $\sum_{i=1}^M x_i'^2\le M/4$, so that this amplification costs $\mathcal{O}(\sqrt{M}\log(M/\epsilon))$, as in the case of uniform sampling.} 
Then we use Lemma~\ref{lemma: theorem56} to transform the diagonal entries according to the function $f$ that we want to evaluate, i.e., obtaining the block encoding of the operator $f(A) =  \sum_{i=1}^M f(x_i') \ket{i}\bra{i} $. In order to enact the importance sampling, we need to inject a factor $1/g(x_i')$ in each term of the sum. For the factor $g(x_i')$, we first define:
$$ \Gamma_i \equiv  \frac{1/g(x_i')}{ \sqrt{ \sum_{i=1}^M (1/g(x_i'))^2 }  } \equiv \frac{\gamma}{g(x_i')}. $$
Then, we can prepare the state $\sum_{i=1}^M \Gamma_i \ket{i}$ using the amplitude encoding method \cite{prakash2014quantum}, and  use Lemma~\ref{lemma: diagonal} to prepare the block encoding of the diagonal matrix:
$$ B = \sum_{i=1}^M  \Gamma_i \ket{i}\bra{i} = \sum_{i=1}^M \frac{\gamma}{g(x_i')} \ket{i}\bra{i}.$$
Thus, we can use Lemma~\ref{lemma: product} to prepare the block encoding of the product $f(A) \cdot B$, which is:
\begin{align}
    f(A) \cdot B = \sum_{i=1}^M \frac{\gamma f(x_i')}{g(x_i')} \ket{i}\bra{i} = \gamma \sum_{i=1}^M \frac{ f(x_i')}{g(x_i')} \ket{i}\bra{i}.
\end{align}
Following the same procedure as in the previous section, we can therefore estimate the summation: 
$$ I = \frac{1}{M} \sum_{i=1}^M \frac{\gamma f(x_i') } { g(x_i')} , $$
which contains our desired integral multiplied by a known factor $\gamma$. Then, we can follow the same routine as before to estimate $I$ to some additive accuracy $\gamma \epsilon$; then, we can obtain the value of our desired integral with accuracy $\epsilon$. We note that we do not use amplification to remove the factor $\gamma$, whose effect is only manifested in the Hadamard test. The total complexity of our quantum implementation of the importance sampling method is
\revise{$$ \mathcal{O}\big(  \sqrt{M} \log(M d \deg(f)/\epsilon) \log(M) t(f) \deg(f)/(\gamma\epsilon) \big), $$}
where again $t(f)$ is the number of terms in $f$ and $\deg(f)$ is the degree of $f$, as in the previous section. \corr{Note that $1/\gamma=(\sum_{i=1}^M g(x_i')^{-2})^{1/2}$ grows with $M$: since $\mathbb{E}_{x\sim g}[g(x)^{-2}]=\int_{\mathscr{D}}dx/g(x)\ge|\mathscr{D}|^2=1$ by the Cauchy--Schwarz inequality, typically $1/\gamma\gtrsim\sqrt{M}$, so that the total complexity is at least of order $M$ (up to the other factors), i.e., this implementation of importance sampling has no speedup with respect to $M$.} \\

\noindent \textit{\textbf{Comparing with classical method:}} The Classical Monte Carlo method requires as many points as possible for higher accuracy. \revise{The classical complexity is apparently $\mathcal{O}( d M \deg(f) \frac{1}{\epsilon^2} )$ in $d$ dimensions.} \corr{Given QRAM access to the $M$ points, our quantum method evaluates their sample mean with a quadratic speedup with respect to the number of points. However, generating the $M$ points classically and loading them into the QRAM already costs $\mathcal{O}(dM)$, so that there is no overall speedup compared with the classical method. Moreover, the QRAM stores $\mathcal{O}(dM)$ numbers, whereas classical Monte Carlo integration can be carried out in a streaming fashion with $\mathcal{O}(d)$ memory; besides the QRAM, our method uses $\mathcal{O}(\min(d,\deg(f))\log(M) + \log(t(f)))$ qubits (see Appendix~\ref{sec: example}).}

\subsubsection{Integration According to Gaussian Quadrature Method}
This method differs from the previous two in that the points are chosen intentionally, and there are weights associated with corresponding points. Here, we limit ourselves to the univariate integral over a single interval. However, it can be extended to the multivariate setting by using a technique called tensor product quadrature~\cite{gerstner2003dimension}. Roughly speaking, this method breaks multivariate integrals into multiple single-variable integrals and carries them out separately before multiplying them together. In the Gaussian quadrature method, the first step is to choose some desired polynomial, e.g., Hermite polynomial, Legendre polynomial, or Chebyshev polynomial with certain degree $n$. Then, we find roots of such polynomials in the interval of integration $(-1,1)$. The integration in such a single interval domain is approximated by:
\begin{align}
    \int_{-1}^{1} f(x) \corr{w(x)}\, dx  \corr{\approx} \sum_{i=\corr{0}}^{\corr{n-1}} w_i f(x_i), 
\end{align}
where \corr{$w(x)$ is the weight function associated with the chosen family of polynomials,} $w_i$ is the weight determined by the choice of the polynomial, and $x_i$ is a node, which is the $i$-th root of the chosen polynomial from this approach. In this case, we choose the Chebyshev polynomial, which can be easily implemented using QSVT~\cite{gilyen2019quantum}. With this choice, the weight function is \corr{$w(x)=(1-x^2)^{-1/2}$, and the quadrature weights are} $w_i = \pi/n$ for all $i$. The roots of the Chebyshev polynomial are given by $x_i = \cos\Big( \frac{(2i+1)\pi}{2n}\Big)$ for $i=0,1,2,...,n-1$.
\begin{corrblock}
To integrate $f$ with unit weight, we apply the rule to $f(x)\sqrt{1-x^2}$:
\begin{align}
\int_{-1}^{1}f(x)\,dx=\int_{-1}^{1}\frac{f(x)\sqrt{1-x^2}}{\sqrt{1-x^2}}\,dx\approx\frac{\pi}{n}\sum_{i=0}^{n-1}f(x_i)\sin\Big(\frac{(2i+1)\pi}{2n}\Big),
\label{eq: gausscheb}
\end{align}
where we used $\sqrt{1-x_i^2}=\sin((2i+1)\pi/(2n))$. (Since $\sqrt{1-x^2}$ is not a polynomial, this rule is not exact for polynomial $f$; its error is $\mathcal{O}(n^{-2})$ for smooth $f$.)
\end{corrblock}
\revise{But Lemma~\ref{lemma: uniformloading} only allows us to block encode  $i$ in the set $[-n/2,-n/2+1,...,n/2-1]$ onto the diagonals of a matrix (up to a scaling factor), which mismatches the  index $i$ in the cosine:  $i\in[0,1,2,\cdots,n-1]$.} \revise{In order to match with the roots of the Chebyshev polynomial, we can adjust the rotational step from Eqn.~(\ref{eqn: rotation}) as following:
\begin{align}
    \ket{0}\ket{i} \longrightarrow  \Bigg(\corr{\frac{2i+1}{4n}}\ket{0} +  \sqrt{1-\Big(\corr{\frac{2i+1}{4n}}\Big)^2}\ket{1}\Bigg) \ket{i}. 
\end{align}
}
Then we follow the same procedure as in the previous section to obtain the $\epsilon$-approximation block encoding of the operator \revise{$\sum_{i=0}^{n-1} \corr{\frac{2i + 1}{4n}} \ket{i}\bra{i}$}, which contains the factor \revise{$\corr{(2i+1)/(4n)}$} on the diagonal. \corr{The factor $1/(4n)$, instead of $1/(2n)$, keeps the diagonal entries below $1/2$, so that the amplification of Lemma~\ref{lemma: amp_amp} can be used with $\delta=1/4$ as in Lemma~\ref{lemma: uniformloading}.}

In order to obtain $f(x_i)$ for $x_i$ being $i$-th root of chosen Chebyshev polynomial, we simply first use QSVT to transform \revise{$\corr{(2i+1)/(4n)} \longrightarrow \cos( \frac{(2i+1)\pi}{2n})$}, which is doable using Lemma (57) of~\cite{gilyen2019quantum}. More specifically, Lemma (57) gives a tool for approximating $\cos (tx)$ \corr{and $\sin(tx)$} (for $x \in (-1,1)$) with polynomial of rough degree $\mathcal{O}(t+ (\log(1/\epsilon)))$. If we choose \revise{$t=\corr{2\pi}$} then we obtain the approximation of $\cos((2i+1)\pi/2n)$ \corr{and, in the same way, of $\sin((2i+1)\pi/2n)$}. From the block encoding of the diagonal matrix that contains $\{\cos( \frac{(2i+1)\pi}{2n}) \}_{i=0}^{n-1}$, we treat them as variables and again use the QSVT (Lemma~\ref{lemma: theorem56}) to transform them to the function $f$ of interest.
\begin{corrblock}
Lemma~\ref{lemma: product} then gives a block encoding of $D=\mathrm{diag}\big(f(x_i)\sin\frac{(2i+1)\pi}{2n}\big)_{i=0}^{n-1}$. The weights need not be inserted into the block encoding: as in Eq.~(\ref{eq: sandwich}), the amplitude $\bra{0}^{\otimes\log n}H^{\otimes\log n}DH^{\otimes\log n}\ket{0}^{\otimes\log n}=\frac1n\sum_iD_{ii}$ already contains the factor $1/n$, so that
\begin{align}
\int_{-1}^{1}f(x)\,dx\approx\pi\,\bra{0}^{\otimes\log n}H^{\otimes\log n}DH^{\otimes\log n}\ket{0}^{\otimes\log n},
\end{align}
where the factor $\pi$ is applied classically and the amplitude is estimated to precision $\epsilon/\pi$.
\end{corrblock}

\smallskip\noindent \textit{\textbf{Comparing with classical method:}} Classically, one needs to evaluate the function $f$ at $n$ roots of the $n$-th degree Chebyshev polynomial, which takes $\mathcal{O}(n T_f)$ time, where $T_f$ is the time required for computing $f$, plus $\mathcal{O}(n)$ time for doing summation. Therefore, in total, it takes $\mathcal{O}(n T_f)$ time steps. 

\revise{Using quantum computers, it first takes $\mathcal{O}( \sqrt{n} \log(\frac{n}{\epsilon})  \log(n) )$ to construct the $\epsilon$-approximated block encoding of diagonal operators that contain \corr{$(2i+1)/(4n)$} for all $i$. Then one use lemma \ref{lemma: qsvt} to transform such diagonal elements to \corr{$\cos( (2i+1)\pi/(2n))$}. We note that, per \cite{gilyen2019quantum}, the cosine function is approximated, for example, with accuracy $\epsilon$ by a polynomial of degree $\mathcal{O}(\log(1/\epsilon))$, therefore, if we input an $\epsilon$-approximated block encoding of \corr{${\rm diag}((2i+1)/(4n))$} to lemma \ref{lemma: qsvt}, then we obtain a \corr{$\mathcal{O}(\log(1/\epsilon) \sqrt{\epsilon})$}-approximated block encoding of \corr{${\rm diag}(\cos((2i+1)\pi/(2n)))$}. The complexity up to this point is $\mathcal{O}( \sqrt{n} \log(\frac{n}{\epsilon})  \log(n) \log(1/\epsilon) )$. Then, we use lemma \ref{lemma: gridpoly} to transform the then $\mathcal{O}(\log(1/\epsilon) \sqrt{\epsilon})$-approximated block encoding of \corr{${\rm diag}(\cos((2i+1)\pi/(2n)))$} to the block encoding of function $f$ of interest. Per Lemma~\ref{lemma: qsvt}, the error approximation of such block encoding is $\mathcal{O}(d \deg(f) \sqrt{\log(1/\epsilon) \sqrt{\epsilon} } )$. We want to obtain an $\epsilon$-approximation (note that we abuse the notation)  eventually, and hence, we need to rescale the error as $d \deg(f) \sqrt{\log(1/\epsilon) \sqrt{\epsilon} } \rightarrow \epsilon $. \corr{The factor $\sin((2i+1)\pi/(2n))$ is constructed in the same way and multiplied using Lemma~\ref{lemma: product}, which changes only constant factors.} Hence, summing up everything, the total time complexity is
$\mathcal{O}( \deg(f) \sqrt{n} \log(n \deg(f)/\epsilon  ) \log(\deg(f)/\epsilon  ) \log(n)  )$ after simplifying powers inside logarithms.} The last step is using amplitude estimation with further $\mathcal{O}(1/\epsilon)$ time. 
Compared to the classical method, the quantum method achieves almost quadratic speedup with respect to $n$. \revise{However, the quantum method also suffers from scaling in the inverse of error dependence and degree of function of consideration. Thus, the almost quadratic speedup is only maintained if the degree of function $f$ is constant $\mathcal{O}(1)$. \corr{The rectangular method has} a better speedup compared to \corr{exact classical summation} in higher dimensions.}
\corr{The quantum method uses $\mathcal{O}(\log(n)+\log(t(f)))$ qubits; note, however, that the classical rule requires only $\mathcal{O}(1)$ memory, since the nodes and weights can be computed on the fly.} % [E15] 

\subsection{Beyond Polynomial Functions}
\label{sec: beyond}
In the above section, we were mainly constructing the variables within a uniformly spaced grid in order to perform numerical integration. The functions we were interested in previously are polynomials, as it is convenient to carry out such transformations using the QSVT framework. One may ask whether we can go beyond polynomials within the QSVT or a more general quantum context. \revise{It turns out one can take advantage of the conditional rotation.}  To proceed, let us look at how we constructed Lemma~\ref{lemma: uniformloading}. We began with the state $\frac{1}{\sqrt{N}} \sum_{j=0}^{N-1} \ket{j}$ created by a single layer of Hadamard gate. In the next part, we performed the following conditional rotation: 
\begin{align}
    \ket{0} \ket{j} \longrightarrow  \Big( f\big( \frac{j}{N}  \big) \ket{0} + \sqrt{1- f^2\big( \frac{j}{N} \big) } \ket{1}   \Big) \ket{j},
\end{align}
\revise{where $f$ is any classically computable function of bounded norm}, e.g., $|f(x)| \leq 1/2$ for $x\in (-1,1)$. The above rotational step is very similar to the way that the HHL algorithm works~\cite{harrow2009quantum}, as after the phase estimation step, one can use such a phase register, which encodes the eigenvalues of a given matrix, to rotate the ancilla and achieve the inversion. On top of that, one can use the phase register to rotate the ancilla with any computable function of desire, and hence, we can perform a transformation of a given matrix beyond just polynomials. 

Therefore, from the above equation, we can follow the same routine as in the previous section, which  applies Lemma~\ref{lemma: diagonal} to obtain the block encoding of 
$$ \frac{1}{\sqrt{N}} \sum_{j=0}^{N-1} f\big(\frac{j}{N}\big) \ket{j}\bra{j}. $$
The factor $\sqrt{N}$ can be removed using the amplification technique as previously. From the block encoding of the above operator, we can estimate the numerical integration of a single-variable function $f$ over the domain $(0,1)$ by noting that:
\begin{align}
    \int_0^1 f(x)dx = \frac{1}{N} \sum_{j=0}^{N-1} f\big( \frac{j}{N} \big).
\end{align}
To deal with multi-variable settings such as a uniformly spaced grid in $d$-dimensions, we perform the following simple adjustment. With $N = 2^n$, we first construct the following state:
\begin{align}
    H^{\otimes dn} \ket{0}^{\otimes dn} = \frac{1}{\sqrt{N^d}} \sum_{j_1,...,j_d=0}^{N-1} \ket{j_1}\ket{j_2} ... \ket{j_d} .
\end{align}
We now use the whole register $\ket{j_1}\ket{j_2} ... \ket{j_d}$ as a controlled register to rotate the ancilla and perform the transformation:
\begin{align}
   \ket{0} \ket{j_1}\ket{j_2} ... \ket{j_d}  \longrightarrow \Big( f\big( \frac{j_1}{N},\frac{j_2}{N},..., \frac{j_d}{N}\big) \ket{0} + \sqrt{1- f^2\big( \frac{j_1}{N},\frac{j_2}{N},..., \frac{j_d}{N}\big) } \ket{1} \Big) \ket{j_1}\ket{j_2} ... \ket{j_d} .
\end{align}
The remaining procedure is the same as the above so that we can estimate the numerical integration of multi-variate function $f$ in a similar manner. The complexity of this method is \corr{$\mathcal{O}( T(f)\, N^{d/2}\, d\log(N/\epsilon)/\epsilon)$, since Lemma~\ref{lemma: diagonal} now yields $N^{-d/2}\,{\rm diag}\big(f(j_1/N,\dots,j_d/N)\big)$, so that the amplification factor is $N^{d/2}$}, where $T(f)$ is the time required to implement the function $f$. It is quite straightforward to see that if $f$ is polynomial, then $T(f)$ is $\mathcal{O}(\deg(f) t(f) )$, where $\deg(f)$ is the degree of $f$ and $t(f)$ is the number of terms in polynomial, and hence\corr{, for $d=1$,} the running time is \corr{comparable to} what we have in Lemma~\ref{lemma: gridpoly}. \revise{Classically, one needs to perform the computation of function $f$ at $N^d$ points, hence taking time complexity $\mathcal{O}(N^d T(f))$. } 
\revise{We summarize various integration methods we have investigated in this work in Table~\ref{table:comparison_integration}.} 

\begin{table}[h!]
    \centering
    \small
    \setlength{\tabcolsep}{3pt}
    \begin{tabular}{|>{\raggedright\arraybackslash}p{0.11\linewidth}|>{\raggedright\arraybackslash}p{0.11\linewidth}|>{\raggedright\arraybackslash}p{0.13\linewidth}|>{\raggedright\arraybackslash}p{0.21\linewidth}|>{\raggedright\arraybackslash}p{0.32\linewidth}|}
    \hline
    method & classical memory size & \begin{tabular}{l}classical\\ runtime
    \end{tabular}
     & quantum memory size & quantum runtime \\ \hline\hline
    rectangular & \corr{$\mathcal{O}(d\log N)$ (streaming)} & $\mathcal{O}(N^d \deg(f) )$ & $\mathcal{O}\big(d\log(N)  + \log(t(f)) \big)$ & $\mathcal{O}\big(N^{1/2}\log\big(N d\, {\rm deg}(f)  /\epsilon\big)$ $\log(N) {\rm deg} (f) t(f)/\epsilon\big)$ \\ \hline
    \begin{tabular}{l}Monte Carlo \\
   uniform and \\
   importance \\
    sampling
    \end{tabular} & \corr{$\mathcal{O}(d)$ (streaming)} & $\mathcal{O}( M \deg(f)  )$ & \corr{$\mathcal{O}\big(\min(d,\deg(f))$ $\log( M ) + \log(t(f))\big)$, plus a QRAM storing $\mathcal{O}(dM)$ numbers} & $ \mathcal{O}\big( \sqrt{ M } \log\big(M d\, {\rm{deg}}(f)/\epsilon\big)$ $\log(M) \deg(f) t(f)/\epsilon \big) $ \corr{given the QRAM (uniform sampling; an additional factor $1/\gamma\gtrsim\sqrt{M}$ for importance sampling), plus $\mathcal{O}(dM)$ for generating and loading the points} \\ \hline
    Gaussian quadrature (1D) & \corr{$\mathcal{O}(\log (n))$} & $\mathcal{O}(n T_f)$ & $\mathcal{O}\big(\log(n) + \log (t (f)) \big)$ &
    $\mathcal{O}\big( \deg(f) \sqrt{n} \log(n \deg(f)/\epsilon  )$ $\log(\deg(f)/\epsilon  ) \log(n)  / \epsilon \big)$ \\ \hline 
    Computable functions & \corr{$\mathcal{O}(d\log N)$ (streaming)}  & $\mathcal{O} (N^d T(f) )$ & $\mathcal{O}( d\log(N) )$ & \corr{$\mathcal{O}\big( T(f)\, N^{d/2}\, d\log (N/\epsilon)/\epsilon \big)$} \\ \hline
    \end{tabular}
    \caption{\revise{Comparison of quantum and classical methods for numerical integration. $N$ in both rows of the rectangular method and the computable functions is the number of points in each of the $d$ directions (or dimensions). $M$ in the second row of the Monte Carlo method represents the total number of points used. $n$ in the row of Gaussian quadrature represents the number of roots used.  } \corr{The classical runtimes refer to exact summation over the given points; for an $\epsilon$-additive estimate on the grid, classical averaging over $\mathcal{O}(1/\epsilon^2)$ random grid points suffices. ``Streaming'' means that the classical sum is accumulated point by point without storing the points. Note that in the Gaussian quadrature, if we use other polynomials, whose roots cannot be computed on the fly, additional memory is needed.}
    }
\label{table:comparison_integration}
\end{table}

\section{Conclusion}\label{sec:conclude}
In this work, we have \corr{studied} two particular problems: the largest eigenvalue estimation and numerical integration based on the quantum singular value transformation framework.
\begin{corrblock}
We specifically show that the matrix powers in the quantum power method proposed in~\cite{nghiem2023quantum} can be implemented with block encodings, which makes the cost of this subroutine polylogarithmic in the dimension and in the inverse error tolerance. The bottleneck of the original quantum power method~\cite{nghiem2023quantum}, i.e., the measurement step whose success probability is exponentially small in the number of iterations $k$, remains: amplitude amplification reduces its cost only quadratically, and, contrary to the claim in the published version of this work, it cannot be removed by a singular value transformation of a block-encoded rank-one operator (see the Remark in Sec.~\ref{sec: acceleratedpower}). Hence, the runtime in Theorem~\ref{thm: main} is still exponential in $k$.
\end{corrblock}

In addition, we also show how to execute several well-known numerical integration techniques, such as the rectangular method, the Monte Carlo method, and the Gaussian quadrature method, with a quantum singular value transformation framework. We have shown how to construct a grid, consisting of equally spaced points with certain coordinates, as a diagonally block-encoded operator, which is inspired by the recent result~\cite{gonzalez2024efficient}. As the next step, we have also shown how to construct a polynomial computed on such a grid, which is carried out naturally as quantum singular value transformation is naturally suited for polynomial transformation. Subsequently, the numerical integration is then evaluated using the well-known Hadamard/SWAP test. For the rectangular method \corr{and the} Gaussian quadrature method, the speedup is polynomial with respect to the number of points\corr{, compared with exact classical summation; for the Monte Carlo method with classically sampled points, the cost of generating and loading the points removes the speedup}.
We do not know whether exponential speedup can be achieved. \corr{Since classical numerical integration can be carried out in a streaming fashion with logarithmic memory, these quantum algorithms do not offer an exponential saving of memory.} 

Along the way, we have provided an alternative solution to the linear loading problem discussed in~\cite{gonzalez2024efficient}. 
Given that QSVT has been shown to be both powerful and enable great flexibility in operations, it is desirable to see if one can improve the speedup further using the versatile tools from quantum singular value transformation~\cite{gilyen2019quantum}. For instance, it is interesting to explore if we can incorporate the idea of our quantum power method with that of~\cite{chen2024quantum} to construct another version of the eigenvalues finding framework. 
As another example, one might ask if our quantum rectangular and Gaussian quadrature methods can be improved to anything better than an almost quadratic speedup \revise{and if a suitable generalization to the multivariate case will have better complexity than its classical counterpart}. Furthermore, given that (numerical) integration appears in many contexts despite not having exponential speedup, our method may be generalized in a broader scope of applications.  
We expect that even more classical computational problems can be improved through the framework of quantum singular value transformation.

\section{Acknowledgements}
%[add this to arxiv comment] A major mistake corrected. We thank Jens Renders for pointing it out to us.
\corr{We thank Jens Renders for pointing out an error in the published version of this manuscript.} 
 This work was supported by the U.S. Department of Energy, Office of Science, Advanced
Scientific Computing Research under Award Number DE-SC-0012704. We also acknowledge the support by a Seed Grant from
Stony Brook University’s Office of the Vice President for Research and by the Center for Distributed Quantum Processing.

\bibliography{ref.bib}{}
\bibliographystyle{unsrt}

\clearpage
\newpage
\onecolumngrid
\appendix

\section{Preliminaries}
\label{sec: prelim}
Here, we summarize the main recipes of our work. We keep the statements brief and precise for simplicity, with their proofs/ constructions referred to in their original works.

\begin{definition}[Block Encoding Unitary]~\cite{low2017optimal, low2019hamiltonian, gilyen2019quantum}
\label{def: blockencode} 
Let $A$ be some Hermitian matrix of size $N \times N$ whose matrix norm $|A| < 1$. Let a unitary $U$ have the following form:
\begin{align*}
    U = \begin{pmatrix}
       A & \cdot \\
       \cdot & \cdot \\
    \end{pmatrix}.
\end{align*}
Then $U$ is said to be an exact block encoding of matrix $A$. Equivalently, we can write:
\begin{align*}
    U = \ket{ \bf{0}}\bra{ \bf{0}} \otimes A + \cdots
\end{align*}
where $\ket{\bf 0}$ refers to the ancilla system required for the block encoding purpose. In the case where the $U$ has the form 
$$ U  =  \ket{ \bf{0}}\bra{ \bf{0}} \otimes \Tilde{A} + \cdots $$
where $|| \Tilde{A} - A || \leq \epsilon$ (with $||.||$ being the matrix norm), then $U$ is said to be an $\epsilon$-approximated block encoding of $A$.
\end{definition}

The above definition has multiple natural corollaries. First, an arbitrary unitary $U$ block encodes itself. Suppose $A$ is block encoded by some matrix $U$. Next, then $A$ can be block encoded in a larger matrix by simply adding any ancilla (supposed to have dimension $m$), then note that $\Ibb_m \otimes U$ contains $A$ in the top-left corner, which is block encoding of $A$ again by definition. Further, it is almost trivial to block encode identity matrix of any dimension. For instance, we consider $\sigma_z \otimes \Ibb_m$ (for any $m$), which contains $\Ibb_m$ in the top-left corner. We further notice that from the above definition, the action of $U$ on some quantum state $\ket{\bf 0}\ket{\phi}$ is:
\begin{align}
    \label{eqn: action}
    U \ket{\bf 0}\ket{\phi} = \ket{\bf 0} A\ket{\phi} + \ket{\rm Garbage},
\end{align}
where $\ket{\rm Garbage }$ is a redundant state that is orthogonal to $\ket{\bf 0} A\ket{\phi}$.

\begin{lemma}[\cite{gilyen2019quantum} \revise{Block Encoding of a Density Matrix}]
\label{lemma: improveddme}
Let $\rho = \Tr_A \ket{\Phi}\bra{\Phi}$, where $\rho \in \mathbb{H}_B$, $\ket{\Phi} \in  \mathbb{H}_A \otimes \mathbb{H}_B$. Given unitary $U$ that generates $\ket{\Phi}$ from $\ket{\bf 0}_A \otimes \ket{\bf 0}_B$, then there exists a highly efficient procedure that constructs an exact unitary block encoding of $\rho$ using $U$ and $U^\dagger$ a single time, respectively.
\end{lemma}

The proof of the above lemma is given in \cite{gilyen2019quantum} (see their Lemma 45). 
\begin{lemma}[\cite{camps2020approximate} \revise{Block Encoding of a Tensor Product}]
\label{lemma: tensorproduct}
    Given the unitary block encoding $\{U_i\}_{i=1}^m$ of multiple operators $\{M_i\}_{i=1}^m$ (assumed to be exact encoding), then, there is a procedure that produces the unitary block encoding operator of $\bigotimes_{i=1}^m M_i$, which requires \revise{parallel single uses} of 
    %each 
    $\{U_i\}_{i=1}^m$ and $\mathcal{O}(1)$ SWAP gates. 
\end{lemma}
The above lemma is a result in \cite{camps2020approximate}.

\begin{lemma}[\cite{childs2012hamiltonian,gilyen2019quantum} Linear combination of block-encoded matrices]
    Given unitary block encoding of multiple operators $\{M_i\}_{i=1}^m$. Then, there is a procedure that produces a unitary block encoding operator of \,$\sum_{i=1}^m \pm M_i/m $ in complexity $\mathcal{O}(m)$, e.g., using block encoding of each operator $M_i$ a single time. 
    \label{lemma: sumencoding}
\end{lemma}

\begin{lemma}[Scaling Block encoding] 
\label{lemma: scale}
    Given a block encoding of some matrix $A$ (as in~\ref{def: blockencode}), then the block encoding of $A/p$ where $p > 1$ can be prepared with an extra $\mathcal{O}(1)$ cost.  
\end{lemma}
To show this, we note that the matrix representation of RY rotational gate is
\begin{align}
   R_Y(\theta) = \begin{pmatrix}
        \cos(\theta/2) & -\sin(\theta/2) \\
        \sin(\theta/2) & \cos(\theta/2) 
    \end{pmatrix}.
\end{align}
If we choose $\theta$ such that $\cos(\theta/2) = 1/p$, then Lemma~\ref{lemma: tensorproduct} allows us to construct block encoding of $R_Y(\theta) \otimes \mathbb{I}_{{\rm dim}(A)}$  (${\rm dim}(A)$ refers to dimension of matirx $A$), which contains the diagonal matrix of size ${\rm dim}(A) \times {\rm dim}(A)$ with entries $1/p$. Then Lemma~\ref{lemma: product} can construct block encoding of $(1/p) \ \mathbb{I}_{{\rm dim}(A)} \cdot A = A/p$.

\section{\revise{Quantum circuit for Lemma~\ref{lemma: diagonal}}}

\revise{Here, we give the quantum circuit that achieves the block encoding as in Lemma~\ref{lemma: diagonal} based on Ref.~\cite{rattew2023non}.
The block encoding of $A$ is achieved separately for its real part $A^{(0)}$ and imaginary part $A^{(1)}$. 
They are block encoded into $U^{(0)}$ and $U^{(1)}$, respectively. Each of them consists of gadgets build upon the state preparation unitary $U_\text{state prep}$ that loads the wave function $|\psi\rangle$.}

\section{\revise{Constructing Multivariate Function }}
\label{sec: example}

\subsection{Example}
Here, we aim to fill the gap to establish a procedure for multi-variable integration, which is also a proof of lemma \ref{lemma: gridpoly}. Suppose that we have some block encoding of the diagonal operator $X_{ij} = x_i \delta_{ij}$ $(i=1,2,...,n)$ that encodes the variable of interest (for example, in Lemma~\ref{lemma: uniformloading}, the block encoded operator $A$ having variables $j/N$ in its diagonals). If we have multiple block encoding of operators $X$ and $Y$ that encode two different variables of interest, then Lemma~\ref{lemma: tensorproduct} allows us to construct the block encoding of a matrix where the diagonal entries are $\{ x_i\cdot y_j \}$ with $i,j=1,2,...,n$.  We can simply use Lemma~\ref{lemma: theorem56} to construct the block encoding of the polynomial of two variables, which is a polynomial of multi-variables. In reality, some multi-variate polynomials may be trickier to construct. Here, we illustrate the idea with an example of a function $f(x,y) = x^3 + x y^2$ and, say, we want to construct such $f$ given that $x$ and $y$ are discretized along the two directions as $x_i, y_j$ for $i,j = 1,2$ for simplicity. In other words, in this case, we have a grid of size $2 \times 2$. 

We note that in this specific setting, where we limit the discretization points to $2$ in each direction and suppose we have a block encoding of 
$$ X = \begin{pmatrix}
    x_1 & 0 \\
    0 & x_2 
\end{pmatrix}, \ Y  = \begin{pmatrix}
    y_1 & 0 \\
    0 & y_2
\end{pmatrix}.  $$
If we apply QSVT to the following (block encoding of) tensor-product matrix:
\begin{align*}
   X \otimes Y = \begin{pmatrix}
        x_1 y_1 & 0 & 0 & 0\\
        0 & x_1 y_2 & 0 & 0 \\
        0 & 0 &  x_2 y_1 & 0 \\
        0 & 0 & 0 & x_2 y_2 \\
    \end{pmatrix},
\end{align*}
then we can only transform to the polynomial of $x_i y_j$, $i,j=1,2$ but cannot achieve the multi-variate function $f(x_i,y_j)$.
To cope with this issue in our example, we simply make the following adjustment to construct the block encoding of a matrix where the diagonal elements are $\{ f(x_i,y_j) \}|_{i,j=1}^2$. First, we use Lemma~\ref{lemma: tensorproduct} to construct the block encoding of:
\begin{align}
    X \otimes \mathbb{I}_2  = \begin{pmatrix}
        x_1 & 0 & 0 & 0 \\
        0 & x_1 & 0 &  0 \\
        0 & 0 & x_2 & 0\\
        0 & 0 & 0 & x_2 
    \end{pmatrix}.
\end{align}
Separately, we use QSVT (or Lemma~\ref{lemma: theorem56} in particular to do the following transformation): 
\begin{align}
Y = \begin{pmatrix}
    y_1 & 0 \\
    0 & y_2 
\end{pmatrix} \longrightarrow  Y' = \begin{pmatrix}
    y_1^2 & 0 \\
    0 & y_2^2 
\end{pmatrix}.
\end{align}
With the above two block encodings, we then use Lemma~\ref{lemma: tensorproduct} to construct the block encoding of 
\begin{align}
    X \otimes Y' = \begin{pmatrix}
        x_1 y_1^2 & 0 & 0 & 0\\
         0 & x_1 y_2^2 & 0 & 0 \\
         0 & 0 &  x_2 y_1^2 & 0\\
         0 & 0 & 0 & x_2 y_2^2 
    \end{pmatrix}.
\end{align}
Using Lemmas~\ref{lemma: theorem56} and~\ref{lemma: sumencoding}, we can construct the block encoding of the following matrix 
\begin{align}
    \frac{1}{2} \big( (X \otimes \mathbb{I}_2)^3 + (X \otimes Y') \big) = \frac{1}{2} \begin{pmatrix}
       x_1^3 + x_1 y_1^2 & 0 & 0 & 0\\
         0 & x_1^3 + x_1 y_2^2 & 0 & 0 \\
         0 & 0 & x_2^3 + x_2 y_1^2 & 0\\
         0 & 0 & 0 &  x_2^3 + x_2 y_2^2 
    \end{pmatrix},
\end{align}
which contains (with a factor 1/2) our function $f$ evaluated at the points in the $2\times 2$ grid.

This factor will contribute to subsequent operations, which will reduce the probability of success in the final result.  Additionally, if we know that the function $f$ is less than $1/2$ on the chosen interval $(-1/2,1/2)$, one may want to improve the factor of 1/2 to 1 using the amplification method~\cite{gilyen2019quantum}. 
We note that for polynomials with general coefficients, we need to construct a general linear combination of block-encoded matrices, for which we can take the following two possible paths:
\begin{itemize}
\item One can use the Linear Combination of Block Encoded Matrices lemma (Lemma 52 of \cite{gilyen2019quantum}), which constructs $\sum^{m}_{j=1} y_j A_j$ from block encodings $U_j$ of $A_j$ and an $\mathcal{O}(\log(m))$-qubit state-preparation unitary that is associated with the vector $y_j$. This procedure adds $\mathcal{O}(\log(m))$ more ancillas compared to $U_j$. 
Since the classical description of the state for $y_j$ is given, we can construct the state-preparation unitary with a circuit of depth $\mathcal{O}(\log(m))$ using the amplitude encoding method. 
Note that $\mathcal{O}(\log(m)) = \mathcal{O}( \log t(f))$, where $t(f)$ is the number of terms in the polynomial.
\item One can rescale the block-encoded matrices according to Lemma~\ref{lemma: scale} with one additional ancilla and a constant depth circuit.
Then one can add multiple encoded matrices with uniform coefficients (the Linear Combination of Block Encoded Matrices lemma with $y_j= {\rm const.}$), in which case the state preparation unitary in the procedure is of constant depth, i.e., a tensor product of the Hadamard gate.
\end{itemize}
The above consideration can be generalized straightforwardly to any polynomial of consideration, which can be very useful and convenient for numerical integration purposes as we will break the given domain into grids.  \\

\revise{Here we provide another example for the Monte Carlo integration, which has a subtlety that is a bit different from the above grid example. Suppose we want to integrate the function \corr{$f(x,y,z) = x^2 y z $} on the domain $[-1/2,1/2] \times [-1/2,1/2] \times [-1/2,1/2]$. Let's say, we randomize $M = 4$ points $(x_1,y_1,z_1), (x_2,y_2,z_2), (x_3,y_3,z_3), (x_4,y_4,z_4)$.  Then we can use the same procedure discussed above to construct the approximated block encoding of the following operator:
\begin{align}
    X = \begin{pmatrix}
    x_1 & 0 & 0 & 0 \\
    0 & x_2 & 0 & 0 \\
    0 & 0 & x_3 & 0\\
    0 & 0 & 0   & x_4
\end{pmatrix}, \ Y  = \begin{pmatrix}
    y_1 & 0 & 0 & 0\\
    0 & y_2 & 0 & 0 \\
    0 & 0 & y_3 & 0\\
    0 & 0  & 0 & y_4
\end{pmatrix}, \ Z = \begin{pmatrix}
     z_1 & 0 & 0 & 0 \\
     0 & z_2 & 0 & 0 \\
     0 & 0 & z_3 & 0 \\
     0 & 0 & 0  & z_4
\end{pmatrix}
\end{align}
using $\mathcal{O}(\log(4) )$ qubits. Then we use lemma \ref{lemma: product} to construct the (approximated) block encoding of:
\begin{align}
    X^2YZ = \begin{pmatrix}
        x_1^2 y_1 z_1 & 0 & 0 & 0 \\
        0 & x_2^2 y_2 z_2 & 0  & 0 \\
        0 & 0 & x_3^2 y_3 z_3 & 0 \\
        0 & 0  & 0 & x_4^2 y_4 z_4 
    \end{pmatrix},
\end{align}
As the given function $f$ is evaluated at those points of interest, then proceed to use the Hadamard test procedure as outlined in section \ref{sec: integration} to numerically perform the integral. The subtlety of the quantum Monte Carlo method is that, for a more general case of $d$ variables, \corr{the system register requires only $\mathcal{O}(\log(M))$ qubits, independently of the dimension $d$, as all coordinates share the sample index and Lemma~\ref{lemma: product} uses the approximated block encodings of the coordinates of the $M$ points in a sequential manner. Since each application of Lemma~\ref{lemma: product} adds the ancillas of the factors (see the footnote to Lemma~\ref{lemma: newmatrixapp}), the total number of qubits is $\mathcal{O}(\min(d,\deg(f))\log(M))$ when the powers of a single coordinate are formed with Lemma~\ref{lemma: theorem56}.}
}

\subsection{Proof of Lemma~\ref{lemma: gridpoly}}
\revise{Here we provide a formal proof of Lemma~\ref{lemma: gridpoly}.}
\par 
\revise{\textit{Proof:} }
\revise{We start with the unitary matrix $\{ U_{X^{(k)}} \}_{k=1,...,d}$, each of which is an $\epsilon$-approximate block encoding of $X^{(k)} = \text{diag}(x^{(k)}_1, ... , x^{(k)}_N)$ that can be achieved with a circuit of depth $\mathcal{O}(N^{1/2} \log(N/\epsilon) \log(N))$ and $\log N+4$ ancillary qubits according to Lemma~\ref{lemma: uniformloading}.
We wish to construct a block encoding of the matrix
\begin{align}
F = \sum_{(p_1, ..., p_d)} C_{(p_1, ..., p_d)} (X^{(1)})^{p_1} \otimes \cdots \otimes (X^{(d)})^{p_d} \, , 
\end{align}
where the sum runs over tuples describing powers $(p_1, ..., p_d) \in \mathbb{Z}^{\otimes d}$ in each term and $C_{(p_1, ..., p_d)}$ are coefficients in the polynomial $f$.
We denote $A_{(p_1, ..., p_d)}:= (X^{(1)})^{p_1} \otimes \cdots \otimes (X^{(d)})^{p_d}$ and $\widetilde{A}_{(p_1, ..., p_d)}:= C_{(p_1, ..., p_d)} (X^{(1)})^{p_1} \otimes \cdots \otimes (X^{(d)})^{p_d}$. }

\revise{The procedure is as follows:
\begin{enumerate}
\item Using the QSVT lemma~(Lemma \ref{lemma: qsvt}), one can construct $U_{(X^{(k)})^{p_k}}$, which is a $\mathcal{O}(p_k \sqrt{\epsilon})$-approximate 
block encoding of $(X^{(k)})^{p_k}$, with $p_k$ uses of $U_{X^{(k)}}$, single use of controlled-$U_{X^{(k)}}$ gate, and $\mathcal{O}( (\log N +5) p_k )$ single- and two-qubit gates, where the number of ancillary qubits increases by 2.
Each of $U_{(X^{(k)})^{p_k}}$ has the runtime complexity $\mathcal{O}(N^{1/2} \log(N/\epsilon) \log(N)p_k )$ 
while it maintains the spatial complexity as asymptotically the same as $U_{X^{(k)}}$, which is $\mathcal{O}(\log N)$.
\corr{Since $x^{p_k}$ has definite parity and $|x^{p_k}|\le1$ on $[-1,1]$, we use the definite-parity version of Lemma~\ref{lemma: qsvt}, so that no factor $1/2$ is incurred for each variable.} % [E18]
\item Next, Lemma~\ref{lemma: tensorproduct} allows us to take tensor products of the matrices over the $d$ variables.
The unitary matrix  $U_{{A}_{(p_1,...,p_d)}}$, which is a $\mathcal{O}(d \text{deg}(f) \sqrt{\epsilon})$-approximate 
block encoding of $A_{(p_1, ..., p_d)}$, can be constructed with parallel single uses of $\{ U_{(X^{(k)})^{p_k}} \}_{k=1,...,d}$ and $\mathcal{O}(1)$ SWAP gates. 
The runtime complexity of each of $\{ U_{A_{(p_1,...,p_d)}} \}$ is bounded by $\mathcal{O}(N^{1/2}\log(N/\epsilon) \log(N) \text{deg} (f))$ while 
the spatial complexity of each unitary is $\mathcal{O}(\log (N) d)$.
\item \corr{To include the coefficients $C_{(p_1, ...,p_d)}$, we use the linear combination of block-encoded matrices with general coefficients (Lemma 52 of~\cite{gilyen2019quantum}; the first path in Appendix~\ref{sec: example}), with a state-preparation unitary on $\mathcal{O}(\log t(f))$ qubits that loads the weights $|C_{(p_1,\dots,p_d)}|/\sum_{\vec p}|C_{\vec p}|$ and absorbs their signs. This yields a block encoding of $F/\sum_{\vec p}|C_{\vec p}|$, which we rescale to a block encoding of $F$ with Lemma~\ref{lemma: scale} (if $\sum_{\vec p}|C_{\vec p}|<1$), since $\sum_{\vec p}|C_{\vec p}|\le1$ for a properly normalized $f$.}
\item \corr{The linear combination uses each $U_{A_{(p_1,\dots,p_d)}}$ a single time.} 
This increases the runtime complexity by the factor of the number of terms $\mathcal{O}(t(f) )$ and the spatial complexity by additional $\mathcal{O}( \log(t(f)) )$ ancillary qubits.
\end{enumerate}
An $\mathcal{O}(d \text{deg}(f)  \sqrt{\epsilon})$-approximate block-encoding of the operator $F$ is then constructed with the runtime complexity $\mathcal{O}(N^{1/2}\log(N/\epsilon) \log(N) \text{deg} (f))$ and the spatial complexity $\mathcal{O}(\log(N)d + \log(t(f)))$.
Redefining the error as $\epsilon \rightarrow (\epsilon/ d \text{deg}(f) )^2 $, we have that an $\epsilon$-approximate block-encoding of the operator $F$ is constructed with the runtime complexity $\mathcal{O}(N^{1/2}\log(N d \text{deg}(f)   /\epsilon) \log(N) \text{deg} (f) t(f))$ and the spatial complexity $\mathcal{O}(\log(N)d + \log(t(f)))$. 
Notice that constant powers inside logarithms are brought to one since they do not make difference asymptotically. (We have used this convention throughout this paper to simplify expressions.) This completes the lemma. $\square$}

\end{document}